\documentstyle[preprint,aps,amsfonts]{revtex}
\begin{document}

\draft

{\tighten
\preprint{\vbox{\hbox{U. of Iowa Preprint}}}

\title{Precise Determination of the Energy Levels of the Anharmonic  \\
  Oscillator from the Quantization of the Angle Variable}

\author{B. Bacus, Y. Meurice, and A. Soemadi}

\address{Department of Physics and Astronomy\\
University of Iowa, Iowa City, Iowa 52246, USA}

\maketitle

\begin{abstract}%
Using an ansatz motivated by the classical form of $e^{i\phi}$, where
$\phi$ is the angle variable, we construct operators which satisfy
the commutation relations of the creation-annihilation operators
for the anharmonic oscillator.
The matrix elements
of these operators
can be expressed in terms of {\it entire} functions in the position
complex plane.
These functions provide solutions of the Ricatti equation associated with
the time-independent Schr\"odinger equation. We relate the normalizability
of the eigenstates to the global properties of the flows of this equation.
These exact results yield approximations which complement the WKB approximation
and allow an arbitrarily precise determination of the energy levels.
We give numerical results for the first 10 levels with 30 digits.
We address the question of the quantum integrability of the system.
\end{abstract}

\newpage

In recent years, many interesting questions regarding the quantum behavior
of systems which classically exhibit sensitive dependence on the
initial conditions have been raised.\cite{chaos}
Classically integrable systems with a perturbation\cite{kam}
provide rich sequences
of transitions to chaos when the parameter controlling
the perturbation
is increased. Recent studies\cite{prl} of
the energy spectra of the quantum version
of such systems have provided unexpected results. Ultimately, the questions
raised in this context may become relevant for quantum field theory on a
space lattice, i.e., a system of coupled anharmonic oscillators.
It would be interesting to understand if the tool which is used classically
to control the effects of the perturbation, namely the analysis of small
denominators,
can also be used for the quantum problem. The prerequisite for such a
discussion is to have at hand a quantum version of the action-angle variables
for classically non-linear problems. Due to the ordering problem
this is a non-trivial issue.

In this letter, we construct a quantum operator corresponding to
the classical quantity
$e^{i\phi}$, $\phi$ being the angle variable, for one of
the simplest non-linear problems with one degree of freedom, namely
the anharmonic oscillator with an energy operator
\begin{equation}
\widehat{H}={\widehat{p}^2\over 2m}+{1\over2} m \omega _o^2
\widehat{x}^2 +\lambda \widehat{x}^4 \ .
\end{equation}
We assume $\lambda >0$, but the sign of $\omega_o^2$ is not crucial for the
calculations which follow. The method proposed can indeed be extended
straighforwardly to the case where $V$ is any even polynomial
bounded from below.
Classically, the angle variable $\phi$ satisfies the Poisson bracket
relation $\{ H, e^{i\phi}\}=-ie^{i\phi}\omega(H)$,
where $\omega $ is one over the derivative of the action with
respect to the energy.
A quantum version of this equation reads
\begin{equation}
[\widehat{H},\widehat{A}]=\hbar \widehat{A} \Delta(
\widehat{H})\ ,\
\end{equation}
where $\Delta $ is a function which remains to be specified.
In order to construct an operator $\widehat{A}$ satisfying Eq. (2), we use
an ansatz which in the classical case provides
an explicit expression for $e^{i\phi}$ in terms of $p$ and $x$.
The form of the classical ansatz can be obtained directly
from the well-known
expression of the angle variable in terms of the position and the energy.
In addition, we pick a special type of ordering, namely
\begin{equation}
\widehat{A}=\sum_{n=0}^{\infty} \widehat{x}^{2n}(i\widehat{p}
K_n(\widehat{H})+\widehat{x}L_n(\widehat{H}))\ .
\end{equation}
With this choice of ordering, a solution $A$ multiplied on the right
by an arbitrary function of $H$ is another solution. In the classical
theory, this ambiguity is raised by imposing that $e^{i\phi}$ be
on the unit circle.
In the case of the potential given in Eq. (1),
we obtain a formal solution of Eq. (2) provided that for $n\geq 0$
\begin{eqnarray}
{\hbar^2\over m} (n+2)&(n+1)(2n+3)K_{n+2}(\widehat{H})=
[-(2\widehat{H}+\hbar\Delta (\widehat{H}))2(n+1)K_{n+1}(\widehat{H})\nonumber
\\
&+((2n+1)m\omega^2_o - {m(\Delta (\widehat{H}))^2
\over{2n+1}})K_n (\widehat{H})
+\lambda 4n K_{n-1}(\widehat{H})] \\
&L_n(\widehat{H})=-{m\Delta (\widehat{H})\over{2n+1}} K_n(\widehat{H}) -\hbar
(n+1) K_{n+1}(\widehat{H})\ . \nonumber
\end{eqnarray}
If $\hbar $ is set to zero in Eq. (4) and $\widehat{H}$ replaced by a
numerical value $E$,
one recovers the
difference equations for the corresponding classical equation
mentioned above.
Our starting Eq. (2) can
be compared with the operator equation
$[F(\widehat{x},\widehat{p}),\widehat{H}]=i\hbar$,
a quantum version of
$\{\phi /\omega(H),H\}=1$,
solved by C. Bender\cite{time} with a different type of ansatz
(which requires
negative powers of $\widehat{p}$ or $\widehat{x}$).

We now study the matrix element
$<x|A|E>=(\hbar Kd/dx +L)<x|E>$, where $L$ and $K$ are short notations for
$\sum_{n=0}^{\infty}L_n(E)x^{2n+1}$ and
$\sum_{n=0}^{\infty}K_n(E)x^{2n}$ respectively.
In the following, the dependence of $L$, $K$ and $\Delta$ on $E$
is implicit and primes denote $x$-derivatives.
Eq. (2) is
satisfied for an arbitrary potential provided that
\begin{eqnarray}
{\hbar\over 2m}K''&+&{L'\over m}+K\Delta =0 \\
{\hbar\over 2m}L''&-&2K'(E-V)+KV'+L\Delta=0 \nonumber \ .
\end{eqnarray}
In the case of the potential given in Eq. (1), these equations can be
solved using the linear recursion relations given by by Eq. (4) with
$\widehat{H}$ replaced by $E$.
As a result, we get
$<x|\widehat{A}|E>$
as a function of $E,\ \Delta(E),
\ K_0$ and $K_1$.
If both $K_0$ and $K_1$ are zero, then $<x|\widehat{A}|E>=0$,
consequently we require
at least one
of them to be non-zero. In order to fix the ambiguity mentioned above,
we shall impose that the first non-zero
$K_n$ to be 1.
One can prove rigorously that
$|K_n|<C_1(C_2 )^n (n!)^{-{2\over 3}}$
with $C_1$ and $C_2$ independent of $n$.
This bound implies that $L$ and $K$ are ${\it entire}$ functions
of the position seen as a complex variable. This
allows us to controllably approximate these
functions in terms of their truncated power series.
Note that in the classical case, the factorial suppression
is absent and the functions $L$ and $K$ have a common finite radius
of convergence which
reflects the existence of turning points.

We are now in position to construct formal solutions of the time-independent
Schr\"odinger equation. First we notice that
\begin{equation}
L^2-\hbar(L'K-K'L)-2mK^2(V-E)=K_0(\hbar^2 K_1+\hbar m
\Delta K_0 +2mK_0 E)\ .
\end{equation}
Recalling Eq. (4), the equality clearly holds at
$x=0$. In addition, Eq. (5) implies
that the derivative of the LHS is zero, consequently it holds for any $x$.
Adjusting the constants $K_0$ and $K_1$ in such a way that the LHS of
Eq. (6) is zero,
and dividing by $K^2$ (temporarily assuming that $K\neq 0$),
we obtain that $L/K$ satisfies the Ricatti equation
\begin{equation}
\hbar({L\over K})'=({L\over K})^2 - 2m(V-E)\ .
\end{equation}
A detailed study of Eq. (5) shows that Eq. (7) is also satisfied near a
zero of $K$. Parametrizing the wavefunction as
\begin{equation}
<x|E>\ \propto \ e^{-{1\over\hbar}\int^x_{x_o} dy {L(y)\over{K(y)}}}\ ,
\end{equation}
we find  that Eq. (7) is the Ricatti form
of the time independent Schr\"odinger equation.
Note that Eq. (8) implies that
$A|E>=0$.
We still need to specify the conditions under which the RHS of Eq. (6) is zero.
After fixing the arbitrariness
in the coefficients as discussed above, there remains two possibilities.
The first one is $K_0=1$  and
$K_1=-{2mK_0\over{\hbar^2}}({\Delta\hbar\over 2}+E)$ which corresponds
to an even eigenfunction $<x|E>$. The second possibility is
$K_0=0$ and $K_1=1$, which corresponds to an odd eigenfunction,
and for which $L/K$ has automatically the
$-\hbar/x$ singularity at the origin.
In both cases, Eq. (4) define uniquely
$L(x)$ and $K(x)$
given $E$ and $\Delta$.
{}From the uniqueness of the solution of the Ricatti equation,
given a condition at $x=0$
($L/K=0$ in the even case and $K/L=0$ in the odd case), $L/K$
is indeed $\Delta $-independent as one can check order
by order in the expansion of $L/K$
near $x=0$. A particularly convenient choice is $\Delta =0$,
because in this case Eq. (5) and (8)
imply that $|<x|E>|^2 \ \propto \ |K(x)|$.
We shall now use the formal solution of the Schr\"odinger equation to find
sharp upper and lower bounds on the energy levels.

Our basic tool to find accurate
upper bounds on the energy levels will be the
theorem proven in Ref. [5] that
for a Sturm-Liouville problem
the $n$-th eigenfunction divides the fundamental domain into $n$ parts
by means of its nodal points.
For the problem discussed here,
the zeros of $<x|E>$ are the poles of $L/K$ which is
seen most easily by picking $\Delta =0$.
Consequently,
if $K$
has more than $n$ zeros at finite $x$ (which are not zeros of $L$),
then $E>E_n$. Furthermore, if $E$ is decreased continuously, the largest zero
of $K$ increases continuously. When $E$ reaches $E_n$, a pair of zeros
disappears
at infinity.
One can then monitor the ``entrance'' of the zeros in
the region $x\leq a$ while
$E$ increases, by finding $E$ such that $K(a,E)=0$. This can be done using
Newton's method with a an appropriate truncation in the expansion of $K$.
One can then check if the existence of the zero can be established
despite the errors due to the truncation (which can be estimated using
the bound mentioned above).

Lower bounds
can be found from the requirement that the wave function
$<x|E>$ is normalizable.
Due to the fact that $<x|E>$ has a definite parity,
we will restrict the discussion to positive $x$ part of the $(x,L/K)$ plane.
This half plane can be divided into a region where $L/K$ increases and a region
where $L/K$ decreases. The boundary between these two regions is
characterized by
$({L\over K})'=0$ which by Eq. (7) implies $({L\over K})=\pm\sqrt{ 2m(V-E)}$.
For this reason
we call this curve the ``WKB curve''. For the potential of Eq. (1),
one finds that if a
trajectory crosses the WKB curve
$L/K$ continues to decrease when $x$ increases. In other words, the
WKB curve is the boundary of a sink.
Using bounds on $(L/K)'$ coming from Eq. (7) we can prove
that whenever the trajectory crosses the WKB curve, $<x|E>$ defined by Eq. (8)
is not normalizable. Consequently, if $<x|E>$ has $n$ nodes and if the
corresponding
$L/K$ ultimately decreases (when $x$ becomes large enough),
then $E<E_n$.
Note that Eq. (7) shows that the poles of $L/K$ can only be simple and have
a residue $-\hbar$. This implies that the Dunham condition\cite{dunham}
used in semiclassical calculations\cite{wkb} is
automatically satisfied. This also implies that for the potential of Eq. (1),
a normalizable wave function
cannot have a zero in the classically forbidden region
(because $L/K$ could never reach
the positive part of the WKB curve for $x$ larger than the location of
the pole).

In summary, when $E$ is sufficiently close to $E_n$  and $x$
sufficiently large, $L/K$ follows closely the trajectory of the positive part
of the WKB curve. When a certain value of $x$ is reached
$L/K$ depends sensitively on small changes in $E$. A small increase in $E$
creates an additional zero of the wavefunction, a small decrease forces $L/K$
to cross the positive part of the WKB curve and to reach the negative
part of it.
This allows us to find sharp bounds on the energy levels.
The only problem which remains is the control of the round-off
errors.
For a usual double precision computation, this is a serious issue, however
since the linear recursion formula of Eq. (4) requires a number of
operations which only
grows linearly with the maximal order calculated, we can use ``slow''
computational
methods involving
a very high precision. This can be implemented, for instance with
Mathematica, using the
instruction ``SetPrecision[...,100]'' for numbers set with a precision of 100
digits. This method has allowed us to obtain the wave function
with very good precision,
at large $x$,
far beyond the
classical turning point, i.e., where the lowest order WKB approximation
works well.

Proceeding this way, we have calculated the first 10
energy levels in the case $m=1/2$, $\omega_0 =2$ and $\lambda=1/10$.
The results are displayed in Table 1 with 30 significant digits.
These numbers have been obtained by keeping 400 terms in the
expansion of $L$ and $K$
and restricting the calculation to the interval $|x|\leq 7.5$
The starting precision was 100 digits.
The difference between the upper and the lower bounds were required to be less
than $10^{-32}$. These calculations have been performed independently
using Mathematica and Maple.
Our numerical results are in agreement with the existing
literature summarized in Ref. [6] and where numbers up to 15
significant digits can be found.
Note that the numerical precision on the lower bounds obtained with
Newton's methods (which can estimated
from the
Mathematica command ``Precision[...]'')
decreases approximately linearly when the level increases as shown
in the last column of Table 1.
It is clear that these round-off errors
are much smaller than the theoretical precision achieved.
More generally, a preliminary analysis
indicates
that the enterprise of calculating a very large number
of levels with a very large
precision does not face prohibitive (i.e, exponential) growth of computer time.
If this is effectively the case, we could say that the quantum anharmonic
oscillator is
``numerically integrable''.

A more satisfactory outcome
would be to find an operator which would be the analog
of the classical action which satisfies the
Poisson bracket relation $\{ I(H),e^{i\phi}\}=-ie^{i\phi}$.
At the quantum level the corresponding relation
implies
an equally spaced spectrum.
Equivalently, if we had an analytical expression for $\Delta(E)$
corresponding to $\widehat{A}$ being the minimal creation operator
(where $E$ is replaced by $E_n$
and $\hbar \Delta (E)$ by $E_{n+1} -E_n$ in the matrix elements),
we could calculate recursively
the energy spectrum. In both cases, it would mean that we would have at hand
an implicit closed
form expression for the energy spectrum.
Despite interesting attempts,\cite{wkb} such an expression has not been found
and discovering it is a challenge for the future.

It is a pleasure to thank T. Allen, A. Bhattacharjee,
C. Bender, W. Klink, Y. Nambu, W. Polyzou, G. Payne, V. Rodgers,
J. Schweitzer, D. Speiser and J. Weyers for valuable
conversations and comments.

}
\vfill
\eject

{{\centerline{\bf Table 1}}
\vskip .3truecm
{\hfill{\vbox{
\offinterlineskip
\halign{\vrule height.4cm depth.2cm\quad\hfil#\quad&
\vrule height.4cm depth.2cm\quad#\quad&
\vrule height.4cm depth.2cm\quad\hfil#\quad
\vrule height.4cm depth.2cm\cr
\noalign{\hrule}
 $n$&\hfil $E_n\ \ \ \ \ \ \ \ \ \ \ \ \ \ \ \ \ \ \ \  $&\hfil$\ $ s.d.\cr
\noalign{\hrule}
 0&1.06528550954371768885709162879&95\cr
 1&3.30687201315291350712812168469&93\cr
 2&5.74795926883356330473350311848&89\cr
 3&8.35267782578575471215525773464&87\cr
 4&11.0985956226330430110864587493&84\cr
 5&13.9699261977427993009734339568&81\cr
 6&16.9547946861441513376926165088&79\cr
 7&20.0438636041884612336414211074&77\cr
 8&23.2295521799392890706470874343&74\cr
 9&26.5055547525366174174695030067&72\cr
\noalign{\hrule}}
}\hfill}}}
\vskip150pt
\centerline{\bf Table Caption}
\noindent
Table 1: The first ten energy levels $E_n$
and the number of (numerically) significant digits of
the upper bound on $E_n$ (s.d)
in the case $\lambda=1/10$, $m=1/2$ and $\omega_0=2$.
\vfil

\end{document}